\documentclass[12pt]{article}
\usepackage{graphicx,amsmath}
\usepackage{units}
\usepackage{color}

\newlength{\dinwidth}
\newlength{\dinmargin}
\def\be{\begin{equation}}   
\def\ee{\end{equation}}  
\def\bea{\begin{eqnarray}}                      
\def\eea{\end{eqnarray}}
\def\ch1{$\chi(1^+)$}
\def\lapproxeq{\lower .7ex\hbox{$\;\stackrel{\textstyle                                                    
<}{\sim}\;$}}                                                    
\def\gapproxeq{\lower .7ex\hbox{$\;\stackrel{\textstyle                                                    
>}{\sim}\;$}}

\begin{document}

\begin{flushright}                                                    
IPPP/16/08  \\
DCPT/16/16 \\                                                    
\today \\                                                    
\end{flushright} 

\vspace*{0.5cm}

\begin{center}

\vspace*{0.5cm}
\begin{center}
{\Large \bf Two scales in Bose-Einstein correlations  }\\

\vspace*{1cm}

V.A. Khoze$^{a,b}$ A.D. Martin$^a$, M.G. Ryskin$^{a,b}$ and V.A. Schegelsky$^{b}$ \\

\vspace*{0.5cm}

$^a$ Institute for Particle Physics Phenomenology, University of Durham, Durham, DH1 3LE \\
$^b$ Petersburg Nuclear Physics Institute, NRC `Kurchatov Institute', Gatchina, St.~Petersburg, 188300, Russia \\

\end{center}

\begin{abstract}
We argue that the secondaries produced in high energy hadron collisions are emitted by small size sources distributed over a much larger area in impact parameter space occupied by the interaction amplitude.  That is, Bose-Einstein correlation of two emitted identical particles should be described by a `two-radii' parametrization ansatz.  We discuss the expected energy, charged multiplicity and transverse momentum of the pair (that is, $\sqrt{s},~N_{\rm ch}, k_t$) behaviour of both the small and large size components.
\end{abstract}

\end{center}
\vspace{0.5cm}

\section{Introduction}
An effective tool to study the space-time structure of the production amplitude is to measure the Bose-Einstein correlations (BEC) between two identical particles produced in the inclusive hadron interaction, see, for example, \cite
{hbt}$-$\cite{Al}. Consider the situation where we have one pion with momentum $p_1$ emitted at point $r_1$ and another identical pion with $p_2$ and $r_2$.  The inclusive cross section for the two identical particles takes the form
\be
\frac{E_1E_2d^2\sigma}{d^3p_1d^3p_2}~=~\frac{1}{2!}|M|^2\langle2+2e^{irQ}\rangle~=~|M|^2~\langle 1+e^{irQ}\rangle,
\ee
where $M$ is the production amplitude, and where 4-vectors $Q=p_2-p_1$ and $r=r_1-r_2$. The $\langle...\rangle$ denote the averaging over $r_1$ and $r_2$,   The $e^{irQ}$ term is due to the permutation of the identical pions; that is, it allows for the pion with $p_2$ to be emitted from the point $r_1$ and simultaneously for $p_1$ from $r_2$.  As a rule the $Q$ dependence of the amplitude $M$ is relatively flat in comparison with the $Q$ dependence of $e^{irQ}$. Thus we are able to evaluate the size of the pion production domain by studying the $Q$ dependence of the whole cross section $d^2\sigma$.

To extract the effect we compare the measured $Q$ spectrum with a similar one but without BEC.  To be precise we form the ratio
\be
R(Q)~=~\frac{dN/dQ~-~dN_{\rm ref}/dQ}{dN_{\rm ref}/dQ}
\label{eq:RQ}
\ee
where $dN/dQ$ is the two pion distribution integrated over all the variables except $Q$, and $dN_{\rm ref}/dQ $
is the distribution expected in a world without BEC.  There are different ways to choose $dN_{\rm ref}/dQ$. We may measure the $\pi^+\pi^-$ distribution for non-identical pions; or we may change the sign of the three momentum of the second pion ${\vec p}_2 \to -{\vec p}_2$; and so on.  None of these approaches compensates for the $Q$ dependence of $M$ completely; but for the conventional `one-radius' fit
\be
R(Q)~=~\lambda e^{-{\bar r} Q}
\label{eq:RQ1}
\ee
 the different values of the mean radius\footnote{The `mean' radius, ${\bar r} $, is such that $e^{-{\bar r} Q}$ approximates the value of $e^{irQ}$ averaged over $r_1$ and $r_2$.}, ${\bar r} $, extracted from the data are close to each other.  Such analyses of high energy proton-proton interactions at the LHC have been performed by ATLAS \cite{ATLAS}, CMS \cite{CMS} and ALICE \cite{ALICE}.

The problem is that the value of $\lambda$ turns out to be less than 1.  In particular, CMS claim the $\lambda=0.62\pm0.01$ \cite{CMS}. On the other hand from (\ref{eq:RQ}) we expect $R(Q=0)=1$. Moreover, it is clear from Fig.~1 of \cite{CMS}, and the analogous plots of other groups, that the fit does not describe the very low Q data points. This indicates that there should be another component of $R(Q)$ with a larger radius populating the region of small $Q$.

In the present paper we argue that the expected structure of the pion emission domain is highly inhomogeneous. We should consider {\it small} size pion sources distributed in the much larger area of the proton-proton interaction. That is, we are led to parametrize $R(Q)$ by two different mean radii.  We explain the physical origin of this situation below.

\section{Mechanisms for multiparticle production}

It was shown long ago that to describe a high energy (say, proton-proton) interaction it is convenient to first select the subset of diagrams which provides the interaction across a large rapidity gap, and whose contribution does not decrease when the rapidity separation increases \cite{Gribov}. The resulting ladder-like set of diagrams forms Reggeon exchange.  In terms of the hadronic degrees of freedom the corresponding subset of diagrams -- called multiperipheral ladder diagrams (Fig.~1(a)) -- was studied first  by Amati el al. \cite{AFS}. In terms of QCD\footnote{For  simple discussions of the transfer from hadronic to QCD ladders see the Appendix of \cite{KMR}, and \cite{Topical}.} they form the BFKL pomeron \cite{BFKL}.  In a general purpose Monte Carlo these are the diagrams for the DGLAP evolution amplitude (Fig.~1(b)).   Recall that it is not sufficient to consider only one ladder.  To describe multiparticle production we have to consider the possibility of the exchange of a few ladders (Reggeons), see Fig. 1(c). In a Monte Carlo this is called the Multiple Interaction (MI) option.

\begin{figure}
\begin{center}
\includegraphics[height=20cm]{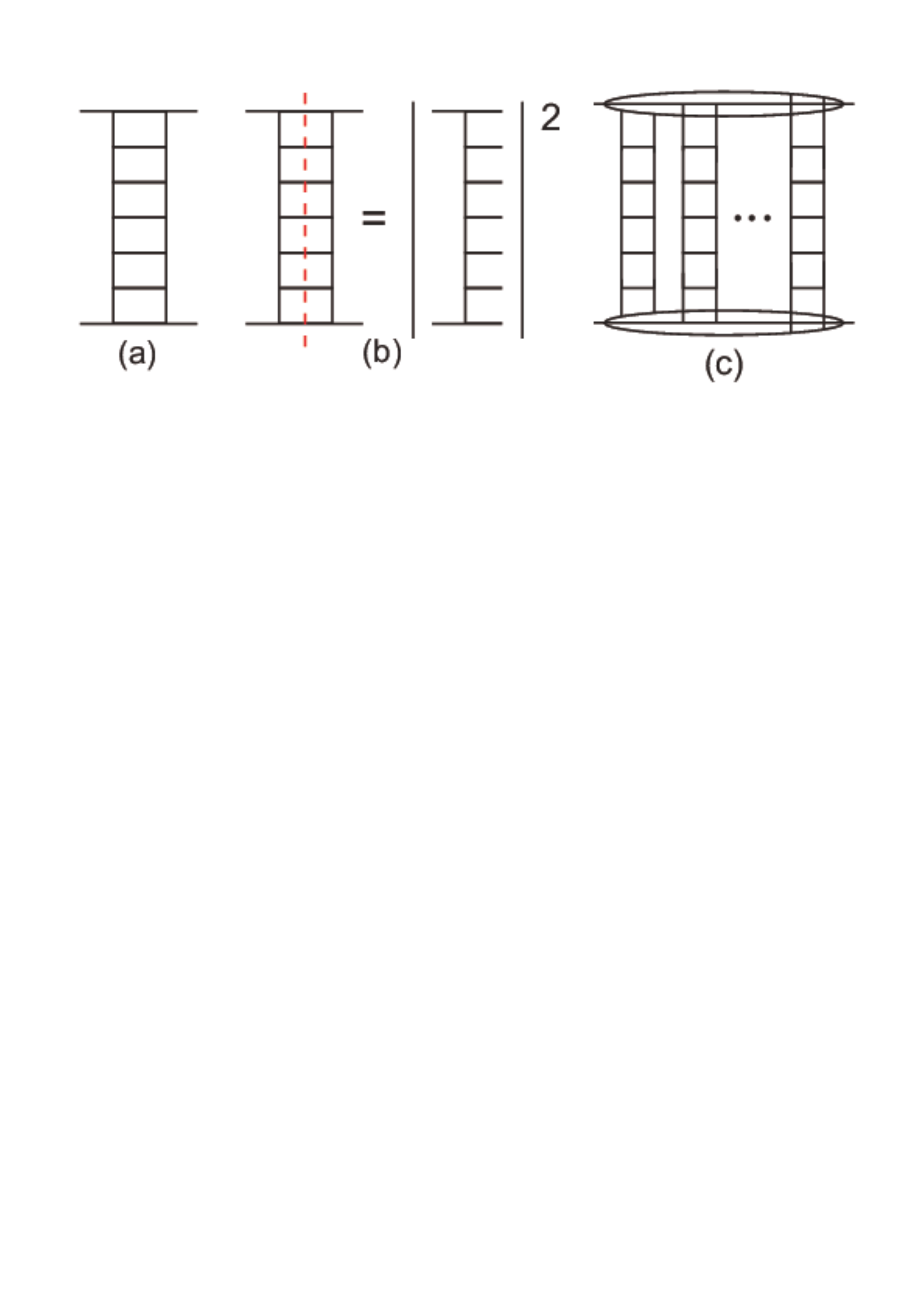}
\vspace*{-10cm}
\vspace*{-4cm}
\caption{\sf (a) The ladder diagram for one-Pomeron exchange; (b) cutting one-Pomeron exchange leads to the multiperipheral chain of final state particles; (c) a multi-Pomeron exchange diagram. }
\label{fig:abc}
\end{center}
\end{figure}

Already at this stage we observe two quite different scales. The slope, $B_{\rm el}$, of elastic proton-proton scattering is usually parametrized in the form
\be
B_{\rm el}(s)~=~B_0+2\alpha'_P{\rm ln}(s/s_0),
\label{eq:slope}
\ee
where the constant $B_0$ is driven by the size of an incoming proton.  On the other hand, the value of $\alpha'_P$ reflects the internal transverse size of the ladder.  Based on the pre-LHC data, typical numbers are $B_0\sim 10$ GeV$^{-2}$ and $\alpha'_P\sim 0.25$ \cite{DL}.

Strictly speaking, (\ref{eq:slope}) corresponds to one-ladder (Reggeon) exchange.  At very high energies the speed of the shrinkage of the diffraction cone increases due to the stronger screening of the amplitude in the centre of the disc, at small impact parameters. Asymptotically, in the black disc limit, where the total cross section grows as ln$^2 s$, the slope $B_{\rm el}(s)$ also increases as ln$^2s$.  It was shown \cite{SChMGR}
that indeed the high energy LHC data indicate the presence of a ln$^2 s$ component in the elastic slope which is consistent with the growth of the total cross section.
However if we consider an individual ladder then the effective value of $\alpha'_P$ in the ladder even decreases with energy due to the larger available $k_t$ space for the intermediate partons.  It is known that $\alpha'_P\to 0$ in the BFKL case.  Another example is the Monte Carlo description of multiparticle production. In order to tune the generator to the high energy data one has to introduce an infrared cutoff, $k_t^{\rm min}$ whose value grows as $s^{0.12}$  \cite{Pythia} -- that is, the transverse size of the ladder decreases.

Thus the interaction of two high energy protons should be described by a diagram like Fig.1(c) in which the size of each individual ladder is rather small (as seen from the small value of $\alpha'_P$). Yet the separation between the ladders is of the order of the radius of the interaction amplitude which should be correlated with the total value of $B_{\rm el}(s)$.   With increasing energy we expect the transverse size of an individual ladder (measured in the central rapidity interval) will decrease (as indicated by the behaviour of $k_t^{\rm min}(s)$).  On the other hand the separation between the ladders is expected to increase, as indicated by the behaviour of $B_{\rm el}(s)$.

\section{Two components in Bose-Einstein correlations}
Having the above picture in mind, we expect in BEC to observe a new object -- a small-size pion source. In other words, BEC should be described by two different radii.
One radius corresponds to the case when both pions are emitted from the same ladder -- this will measure the size of an individual `pomeron'. Since the pion is not a point-like object the radius will be smeared out by fluctuations in the process of the formation of the pions.  The second radius will correspond to the pions being emitted from two different ladders -- it is a measure of the separation between the ladders.
Therefore we propose to fit the observed correlation $R(Q)$ by a formula with two different mean radii\footnote{Instead of the linear exponents, as in (\ref{eq:RQ2}), other parametric forms may be considered for each term. For example, the second term may be a Gaussian, $e^{-({\bar r_2} Q)^2}$. The choice should be based on statistical criteria or on the relative strength of the two terms.   }
\be
R(Q)~=~\lambda~e^{-{\bar r}_1 Q} + (1-\lambda)~e^{-{\bar r_2} Q} ,
\label{eq:RQ2}
\ee
which better reflects the complicated structure of the pion emission domain.
In the ideal case we expect the low multiplicity events to be produced via a diagram with only one `cut' pomeron exchange (Fig.~1(b)).  In general there may be more pomerons in the whole amplitude, but only one ladder radiates the secondaries.  As the multiplicity becomes larger the secondaries are mainly emitted from a few different ladders, and the probability to find the two identical pions originating from the same pomeron decreases.  That is, we expect the relative contribution $\lambda$ of the large component (described by, say, ${\bar r_1}$) to increase with $N_{\rm ch}$, while on the other hand, the strength of the small size component $(1-\lambda)$ decreases.
Unfortunately we cannot predict that $\lambda \to 0$ as $N_{\rm ch}\to 0$, and that $\lambda \to 1$ for very large $N_{\rm ch}$.
The situation is complicated by the strong fluctuations of the number of secondaries in each ladder (or pomeron). Recall that actually we do not measure the total charged multiplicity of an event, but rather the number of secondaries in a limited central rapidity interval (like $|\eta|<2.5$ in the case of ATLAS and CMS). Then the multiplicity corresponding to one pomeron is relatively small ($N_{\rm ch}\sim 4$) and the fluctuations strongly wash out the relation between the measured values and the number of cut pomerons.
Moreover for very low $N_{\rm ch}$ we may sample contributions from diffractive dissociation which have a qualitatively different structure.  Nevertheless at large $N_{\rm ch}$ the multipomeron contribution dominates; that is BEC are driven mainly by the component with the largest radius, ${\bar r}_1$. Indeed it is seen in Fig.~3(b) of \cite{ATLAS} that, in the {\it one} radius fit, (\ref{eq:RQ1}), the radius increases with multiplicity reaching {\it saturation} of ${\bar r}\simeq 2$fm for $N_{\rm ch}\gapproxeq 50$.  

Such saturation was predicted in \cite{SKMR}, where it was explained that the radius measured in a one-radius fit, is driven, not by the initial energy, but mainly by the number of cut pomerons, $n_P \propto N_{\rm ch}$. Indeed, the radius of the individual 
pomeron depends weakly on energy ($\alpha'_P\simeq$ const.), 
while the number of cut pomerons observed in the event (that is the separated pion sources) is 
proportional to $N_{\rm  ch}$. So the probability to have two identical pions from two different sources increases with $N_{\rm ch}$.
When $n_p=1$ (at low $N_{\rm ch}$) we observe one pomeron and measure its radius. On the other hand, for large $N_{\rm ch}$ we 
study the separation between the pomerons and the value of ${\bar r}$ is saturated at the radius of the interaction amplitude ${\bar r} \propto \sqrt{B_{\rm el}(s)}$. It was shown in \cite{SKMR} that $B_{\rm el}\sim 20$ GeV$^{-2}$ \cite{Totem} corresponds in a `one-radius' fit to ${\bar r}=2.2$fm, which is in good agreement with Fig.~3(b) of the ATLAS fit \cite{ATLAS}. 

We emphasize that the separation between the pomerons (the pion sources) is not equal to the incoming hadron (proton) radius. First note, the radius corresponding to the interaction amplitude is larger. It increases with energy.  Recall that in each successive step of the ladder in Fig.~1 
the impact parameter changes by $\Delta b_t$. This leads to a diffusion in $b_t$, which results in the second (or $\alpha'_p$) term in the equation for the elastic slope, (\ref{eq:slope}).  Next, the pion is not a point-like particle and its formation also occupies some volume.  Finally, there may be an interaction between the final state secondaries.

Strictly speaking the picture that we describe above corresponds to the initial stage of the interaction, and does not allow for possible final state rescattering. If there are final state interactions (either in terms of a hadron gas or a quark-gluon plasma) then BEC will measure the radius given by the point of the last interaction -- that is, the domain occupied by secondaries is extended up to the stage where the particle density becomes so low that further interaction is very unlikely, and the size of the domain at which the last interaction occurs will be seen in BEC.   The probability of final state rescattering increases with particle density $N_{\rm ch}$. If this contribution were to dominate, then it is natural to expect ${\bar r} \propto (N_{\rm ch})^{1/3}$. However, the data of Fig. 3(b) of \cite{ATLAS} do not show any evidence of such a behaviour for large $N_{\rm ch}$. 

So what are the expectations of the `two-radii' fit?  The contribution of small raduis dominates at low $N_{\rm ch}$ and decreases with increasing $N_{\rm ch}$ \cite{SKMR}.  The value of the small radius, ${\bar r}_2$, is almost independent of energy -- there is a small tendency to decrease due to a larger $k_t$ in the ladder (see, for example \cite{KMRkt}).  On the other hand, the strength of the ${\bar r}_1$ component increases with $N_{\rm ch}$ -- the value of the radius, ${\bar r}_1$, correlated with $B_{\rm el}$ (that is, the radius of interaction of the incoming protons) slowly increases with energy. At asymptotic energies we expect 
$B_{\rm el} \propto {\rm ln}^2 s$. A tendency already seen in LHC data \cite{SChMGR}. Therefore for $s\to \infty$ we expect ${\bar r}_1 \propto c~$ln$s$. However, the coefficient $c$ is numerically quite small.

For very large $N_{\rm ch}$ the value or ${\bar r}_1$ may additionally increase due to the final state rescattering, as was discussed above. However it is not seen in the 7 TeV $pp$ collision data.

\section{The $k_t$ dependence of BEC}
Here we discuss the dependence of the radii, ${\bar r}_1,~{\bar r}_2$, on the transverse momentum of the identical pair, $k_t=(p_1+p_2)_t/2$,   First, we make a trivial remark --- for larger $k_t$ we have better space resolution in implementing the BEC method.  At low $k_t$ it may be hard to distinguish between ${\bar r}_1$ and ${\bar r}_2$ components, since the radii ${\bar r}_1$ and ${\bar r}_2$ will be larger than that measured at large $k_t$ due to uncertainty principle smearing, and thus closer to each other.

Recall that the probability to produce two large $k_t$ pions from differents ladders (pomerons) decreases with increasing $k_t$ as the single particle inclusive cross section decreases steeply with $k_t$. That is, two identical pions, each with large $k_t$, should be produced from the same large $k_t$ jet.  So with increasing $k_t$ of the pair we expect a large contribution from the ${\bar r}_2$ component with the value of ${\bar r}_2$ decreasing, reaching saturation corresponding to the jet size. To be more precise, we mean the size of pion formation due to the hadronization of large $k_t$ jets. This tendency of ${\bar r}$ to decrease with increasing $k_t$ was indeed observed in a `one-radii' fit of LHC data, see Figs. ~5,6 of \cite{ATLAS}.  However, the value of $k_t$ was not sufficiently large to see saturation in these plots.

Another effect at very large $k_t$, which may give an important contribution to large $N_{\rm ch}$ events, concerns the multiplicity of jets which increase as exp$(-c\sqrt{{\rm ln}E_T})$, see, for example, ~\cite{NE}.  However the high $E_T$ jet cross section is too small to identify this effect in the present data.

\section{Conclusions}
We emphasize that the dynamics of high energy hadron interactions is based on subamplitudes of small transverse size which are distributed over the whole domain occupied by the full interaction amplitude. Thus, in BEC we have to observe a small size object corresponding to the emission of both pions (or both kaons etc.) from a single subamplitude and a larger radius caused by events where the pions are produced from different subamplitudes.  At large $N_{\rm ch}$ the relative contribution  of the large radius component increase.  Enlarging the $k_t$ of the identical pion pair we improve the space resolution of the BEC analyzer. This allows a better separation of the contributions from the small and large radii components. When $k_t$ becomes too large (say, $k_t\gapproxeq1$ GeV) the probability to produce such large $k_t$ pions from two different subamplitudes becomes small.  In this case BEC measure the radius of the `jet' which emits this high $k_t$ pair of identical pions. That is, we expect the radius to decrease with increasing $k_t$, reaching {\it saturation} for $k_t \gapproxeq 1$ GeV.

\section*{Acknowledgement}
MGR was supported by the RSCF grant 14-22-00281. VAK thanks the Leverhulme Trust for an Emeritus Fellowship. MGR thanks the IPPP at Durham University for hospitality.


\thebibliography{}

\bibitem{hbt} R. Hanbury-Brown and R.W. Twiss, Phil. Mag. {\bf 45}, 663 (1954); Proc. Roy. Soc. {\bf 242A}, 300 (1957); {\it ibid} {\bf 243A}, 291 (1957).
\bibitem{gfg} G. Goldhaber, W.B. Fowler, S. Goldhaber et al.,
 Phys. Rev. Lett. {\bf 3}, 181 (1959).
\bibitem{KP} G.I. Kopylov and M.I. Podgoretskii, Sov. J. Nucl. Phys. {\bf 15}, 219 (1972); {\bf 18} 336 (1973).
\bibitem{Al} G. Alexander, Rep. Prog. Phys. {\bf 66}, 481 (2003).
\bibitem {ATLAS} ATLAS Collaboration, G. Aad et al., Eur. Phys. J. {\bf C75} (2015) 466.

\bibitem {CMS} CMS Collaboration, V. Khachatryan et al., JHEP {\bf 1105} (2011) 029.

\bibitem {ALICE} ALICE Collaboration, K. Aamodiet et al., Phys. Rev. {\bf D84} (2011) 112004.

\bibitem{Gribov} V.~N.~Gribov,
 Sov.\ Phys.\ JETP {\bf 26} (1968) 414
 [Zh.\ Eksp.\ Teor.\ Fiz.\ {\bf 53} (1967) 654].


\bibitem{AFS} D.~Amati, A.~Stanghellini and S.~Fubini,
 Nuovo Cim.\ {\bf 26}, 896 (1962).


\bibitem{KMR} 
V.A.~Khoze, A.D.~Martin and M.G.~Ryskin,
 Int.\ J.\ Mod.\ Phys.\ A {\bf 30} (2015) 08,� 1542004
 [arXiv:1402.2778 [hep-ph]].

\bibitem{Topical} V.A.~Khoze, A.D.~Martin, M.G.~Ryskin and A.G. Shuvaev, J.Phys. {\bf G36} (2009) 093001.

\bibitem{BFKL} B.~L.~Ioffe, V.~S.~Fadin and L.~N.~Lipatov,
 ``Quantum chromodynamics: Perturbative and nonperturbative aspects,''
 (Cambridge, UK: Univ. Pr. (2014) 585 p).


\bibitem{DL} 
 A.~Donnachie and P.~V.~Landshoff,
Nucl.\ Phys.\ B {\bf 231}, 231 (1984),
 Nucl.\ Phys.\ B {\bf 244}, 322 (1984).


\bibitem{SChMGR} 
 V.~A.~Schegelsky and M.~G.~Ryskin,
 Phys.\ Rev.\ D {\bf 85} (2012) 094024
 [arXiv:1112.3243 [hep-ph]].


\bibitem{Pythia}
T.~Sjostrand, S.~Mrenna and P.~Z.~Skands,
 Comput.\ Phys.\ Commun. {\bf 178}, 852 (2008)
 [arXiv:0710.3820 [hep-ph]].


\bibitem{SKMR} 
V.~A.~Schegelsky, A.~D.~Martin, M.~G.~Ryskin and V.~A.~Khoze,
 Phys.\ Lett.\ B {\bf 703}, 288 (2011)
 [arXiv:1101.5520 [hep-ph]].


\bibitem{Totem} 
G.~Antchev {\it et al.} [TOTEM Collaboration],
 Europhys.\ Lett.\ {\bf 101}, 21002 (2013).


\bibitem{KMRkt} 
V.~A.~Khoze, A.~D.~Martin and M.~G.~Ryskin,
 Eur.\ Phys.\ J.\ C {\bf 74}, no. 12, 3199 (2014)
 [arXiv:1409.8451 [hep-ph]].


\bibitem{NE}   L.~V.~Gribov, E.~M.~Levin and M.~G.~Ryskin,
Phys.\ Rept.\  {\bf 100} (1983) 1;\\
  V.~A.~Khoze and W.~Ochs,
Int.\ J.\ Mod.\ Phys.\ A {\bf 12}, 2949 (1997)
[hep-ph/9701421].
\end{document}